# CFAAR: Control Flow Alteration to Assist Repair

Chadi Trad, Rawad Abou Assi, Wes Masri, and Fadi Zaraket
Electrical and Computer Engineering Dept.
American University of Beirut
cht02@aub.edu.lb, rawad84@gmail.com, {wm13, fz11}@aub.edu.lb

*Abstract*— We present *CFAAR*, a program repair assistance technique that operates by selectively altering the outcome of suspicious predicates in order to yield expected behavior. *CFAAR* is applicable to defects that are repairable by negating predicates under specific conditions.

*CFAAR* proceeds as follows: 1) It identifies predicates such that negating them at given instances would make the failing tests exhibit correct behavior. 2) For each candidate predicate, it uses the program's state information to build a classifier that dictates when the predicate should be negated. 3) For each classifier, it leverages a *Decision Tree* to synthesize a patch to be presented to the developer.

We evaluated our toolset using 149 defects from the *IntroClass* and *Siemens* benchmarks. *CFAAR* identified 91 potential candidate defects and generated plausible patches for 41 of them. Twelve of the patches are believed to be correct, whereas the rest provide repair assistance to the developer.

## I. INTRODUCTION

Once a failure is detected, it is typically handed over to the developers in order to initiate the debugging process that involves: 1) identifying what caused it, and 2) modifying the code to prevent it from recurring. Researcher working on automating the debugging process refer to the first activity as *fault localization*, and the second as *program repair*. For over three decades, researchers have proposed a plethora of automated fault localization techniques and tools [7][10][15][20][21][33][39][43][35]. And in recent years a number of automated program repair techniques have been proposed that leverage varying approaches such as evolutionary algorithms [19][37], constraint solving [12][14][9][26][27], and program mutation [8]. Long and Rinard [24], Xiong et al. [40], Demarco et al. [9], and Xuan et al. [41] proposed repair techniques that are focused on condition synthesis, which pertains most to our work.

We present *CFAAR* (*Control Flow Alteration to Assist Repair*), a test-based program repair assistance technique that operates by selectively altering the outcome of suspicious predicates in order to yield expected behavior, and subsequently provide a synthesized patch. It focuses on the category of defects that are repairable by *negating* control statements under some specific conditions. Unlike most other test-based repair techniques that mine for patches in other parts of the program [17][18] or in various artifacts, *CFAAR* relies on the program's state to determine when a candidate control statement should be negated in order to yield correct behavior. The captured state information is further analyzed in order to synthesize a patch in the form of a conditional that guards the candidate control statement. When presented with a patch, the developer would: 1) use it as is, if deemed correct; or 2) use it as assistance during the debugging process.

Specifically, given a test suite in which the test cases are classified as failing or passing, *CFAAR* operates as follows:

> *Step1*. It identifies a set of suspicious predicates using an existing coverage-based fault localization (CBFL) technique.
>
> *Step2*. For each suspicious predicate, it uses a heuristic search to identify execution instances such that negating the predicates at the given instances would make some (but not necessarily all) of the failing tests exhibit correct behavior. Our repair assistance approach would be deemed to have failed in case this step was unable to make any failing test case exhibit correct behavior.
>
> *Step3*. For each candidate predicate, a classifier is built whose purpose is to dictate when the predicate should be negated to yield correct behavior. The training output data for the classifier (to negate vs. not to negate) is deduced from the execution instances identified in *Step2*. The training input data is derived from the program state captured at the point of predicate execution. It is worth pointing out that, in several cases we encountered, if *Step2* was only able to make part of the failing tests exhibit correct behavior, the built classifier might compensate for that shortcoming, as discussed in Section II.*C*.
>
> *Step4*. For each classifier, *CFAAR* leverages a *Decision Tree* in order to synthesize a corresponding patch deployable in the form of a conditional statement guarding the candidate predicate. The developer might deem the patch correct and adopt it as a fix, or simply use it to guide and assist the debugging process.

The main contributions of this work are:

a. A program repair assistance approach that is centered on selectively altering control flow, with specific focus on defects that are repairable by negating control statements under specific conditions.
b. A supporting toolset that targets the Java platform.
c. An evaluation of the toolset demonstrating its effectiveness at generating synthesized patches for the *Introclass* benchmark and part of the *Siemens* benchmark.

Section II provides a detailed description of our repair assistance approach. In Section III, *CFAAR* is evaluated by applying it on the *Introclass*, and *Siemens* benchmarks. Section IV surveys related work, and Section V concludes.

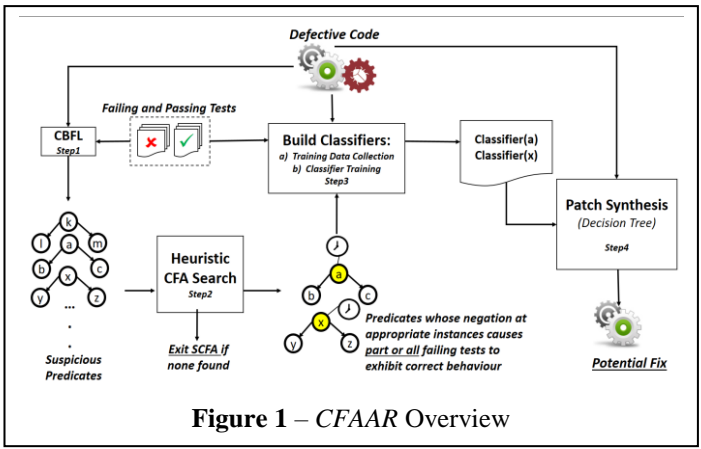

**Figure 1** – *CFAAR* Overview

## II. CFAAR: SELECTIVE CONTROL FLOW ALTERATION

Our program repair assistance approach is based on the premise that a measurable proportion of the defects are likely to trigger erroneous branch executions. As such, we expect that properly altering suspicious predicates at runtime is likely to cause the failure to disappear, and thus enable us to synthesize a potential code fix. This section describes *CFAAR* in detail.

### A. Overview

Figure 1 provides an overview of our approach. Given a set of passing test cases and another of failing test cases, a CBFL technique is applied whose outcome is a set of suspicious predicates.

A *Heuristic Search for Control Flow Alterations* (see Section II.*B*) is applied on the suspicious predicates to identify a minimal set of predicates whose negation at proper instances causes the failure to disappear in some or all test cases that were originally failing. *CFAAR* is deemed unsuccessful in case the search failed to find any candidate predicates and associated execution instances.

The *Build Classifiers* phase creates for each candidate predicate $p$ a classifier whose purpose is to dictate when the predicate should be negated to yield correct behavior. This involves two steps:

 *a) Training Data Collection*: the test suite is executed in order to capture the program states relevant to $p$, every time $p$ is executed. The captured states are labeled as those for when $p$ needs to be negated and those for when a negation is not required.

 *b) Classifier Training:* involves using the collected states to train a *Decision Tree* classifier that decides whether or not to negate $p$.

The *Patch Synthesis* phase generates a potential code fix by: a) building a *Decision Tree* out of each classifier; and b) converting each tree into a predicate that will guarantee that the corresponding suspicious candidate predicate is negated appropriately. That is, each synthesized patch should faithfully replace its corresponding classifier.

### B. Heuristic CFA Search

We devised *HeuristicCFASearch*, a search algorithm that identifies which predicates to negate and when to negate them. Specifically, the goal is to identify according to which pattern of execution, picked from the list of pre-determined patterns shown in Figure 2, a certain predicate should be negated. For example, the "all" pattern means that the predicate should be negated all the time, and the "first+last" patterns means that it should be negated only the first time it executes and the last time.

Note how each of the supported patterns of execution are generic enough to be matched across different test runs. Consider for example a predicate $p$ that executes 7 times in failing test case $t_1$ and 8 times in $t_2$. Now assume that we discover that $t_1$ passes if $p$ is negated according to the pattern "0111110". Finding a matching pattern in $t_2$ is possible and yields "01111110" according to the generic pattern "all-(first+last)". However, if the discovered pattern was "1101011" it is not possible to find a unique match in $t_2$. For that reason, we restrict our search to patterns that are easily reproducible across different execution runs.

*HeuristicCFASearch* considers a single suspicious predicate $p$ at a time and a set of different patterns of negation. Specifically, it checks whether any of the following actions would make some or all of the failing test cases succeed: 1) negating $p$ all the time within a given failing test case; 2) negating $p$ the first time; 3) negating $p$ the last time; 4) negating $p$ all the time except the first; 5) negating $p$ all the time except the last; and so on, as indicated on Line 1 of the pseudocode shown in Figure 2.

*HeuristicCFASearch* takes as input: 1) *PredList$_{susp}$*: the list of suspicious predicates identified by the CBFL component; and 2) $T_{fail}$: the set of failing test cases within the training set. Line 1 initializes *Patterns* with the execution patterns to be matched; note that the patterns are roughly ordered in terms of their simplicity. On Line 2, *PredList$_{solution}$* is initialized to the empty set; its role is to store the suspicious predicates that are candidates for repairing one or more failing runs. For every suspicious predicate $p$, every failing test $t_{fail}$, and every pattern *pattern*, Line 6 executes $t_{fail}$ while negating $p$ according to *pattern*. In case the execution succeeds, $p$ is deemed to be a viable candidate for repairing $t_{fail}$ according to *pattern*. Accordingly, Line 8 associates $p$ with $t_{fail}$ and *pattern*, and Line 9 adds $p$ to *PredList$_{solution}$*.

Lines 10-13 orders all of the (*p, pattern*) pairs based on the number of failing test cases they fixed. The ordered list is stored in the priority queue *PredPatternPQ$_{solution}$*, and returned at Line 14.

### C. Building the Classifiers

*Training Data Collection* – We train one classifier at a time for every pair (*p, pattern*) in the ordered list identified by *HeuristicCFASearch*. The objective is to obtain one or more classifiers that can *plausibly* fix the subject program. If multiple classifiers turn out to be plausible, the one that facilitates synthesis the most will be considered.

```
HeuristicCFASearch(PredList_susp, T_fail)

1. Patterns = {"all", "first", "last", "all-first", "all-last",
   "all-(first+last)", "first+1", "last-1", "first+last", "odd", "even"}

2. PredList_solution = ∅

3. for each p in PredList_susp  do
4.    for each t_fail in T_fail do
5.       for each pattern in Patterns do
6.          execute t_fail while negating p according to pattern
7.          if execution succeeds
8.             p.repairs(t_fail, pattern)
9.             PredList_solution = PredList_solution U p
             endif
          endfor
       endfor
    endfor
endfor

// order the (p, pattern) pairs by the number of the test cases they fix

PredPatternPQ_solution = ∅          // priority queue

10. for each p in PredList_solution do
11.    for each pattern in Patterns do
12.       priority = p.getNumFixedTCsByPattern(pattern)
13.       PredPatternPQ_solution.insert(priority, (p, pattern))
       endfor
    endfor

14. return PredPatternPQ_solution
```

**Figure 2** – Heuristic CFA Search Algorithm

Given a pair (*p, pattern*) we collect data to train a classifier that will guide the execution by indicating when to negate *p* according to *pattern*. Two sets of data are actually needed, one associated with when *p* needs to be negated and another associated with when *p* should remain intact. In other words, we need to capture the states induced by: 1) the failing test cases that were fixed using (*p, pattern*); and 2) all the passing test cases.

The two sets are built by collecting the approximated state of the program right before each execution of *p*. Specifically, on the onset of *p* executing, the values of the following entities are collected:

1) *Use*(*p*), i.e., the local variables, static variables, array elements, instance fields, and method return values, directly *used* in *p*.

2) Formal parameters of the method containing *p*.

3) Local and static variables that were *used* or *defined* within the method containing *p*.

The values of the program variables are derived according to their types, as follows:

1) Variables of type *float* and *double* have their values used as is, i.e., as scalars.

2) Variables of type *int*, *long*, *char*, *byte*, and *short* have their values used in a dual manner, as scalars and categorical.

3) *java.lang.String* objects are categorically represented within the classifier, such that the categories are determined based on the *java.lang.String.hashCode()* method.

4) Non-*String* objects are also categorically represented within the classifier. However, the categories are determined based on a hash code computed by considering the states of the objects' attributes, and if need be, by recursively considering the attributes of their attributes and so on. In other words, to represent a non-*String* object, a hash code is first derived based on all of its direct and indirect attributes.

*Training the Classifiers* – In this phase, a classifier is trained using the previously collected training data. The outcome is a *Decision Tree* that tests one variable at a time to determine if the candidate predicate should be negated. Since the operations that process decision trees are greedy by nature, we expect to have a small number of variables in the tree and consequently only few variables in the resulting synthesized patch.

*D. Patch Synthesis*

Given a classifier that makes all test cases pass, the **Patch Synthesis** phase generates a synthesized patch by building a *Decision Tree* out of the classifier, and then converting it into a predicate that guards the suspicious candidate predicate. This process is detailed below and illustrated in Figure 3.

The computed decision tree serves as a blueprint of the patch. Its leaves indicate whether the predicate should be negated or not. The tree is first converted to a Boolean expression as follows (see Figure 3):

1) Every path from the root node to a leaf indicating a negation is transformed into a rule comprising a conjunction of the conditions along the path.

2) The obtained rules are grouped using disjunctions.

3) The final expression is a DNF formula consisting of literal equalities and inequalities. We note that the generated expression can be reduced through various methods, but this doesn't affect the correctness of the expression.

The resulting Boolean expression is transformed into bytecode following the steps below:

1) The variables used in the patch are identified. Local variables and static fields can be used directly in the patch. However, instance fields and method returns cannot; they are stored in temporary local variables that the patch will use.

2) Compatible bytecode operators for the equalities and inequalities in the Boolean expression are identified. For example, a variable of type *int* will require a different comparison operator than what a variable of type *double* would require.

3) The Boolean expression is transformed into a sequence of *if* statements that determine if the condition is met or not.

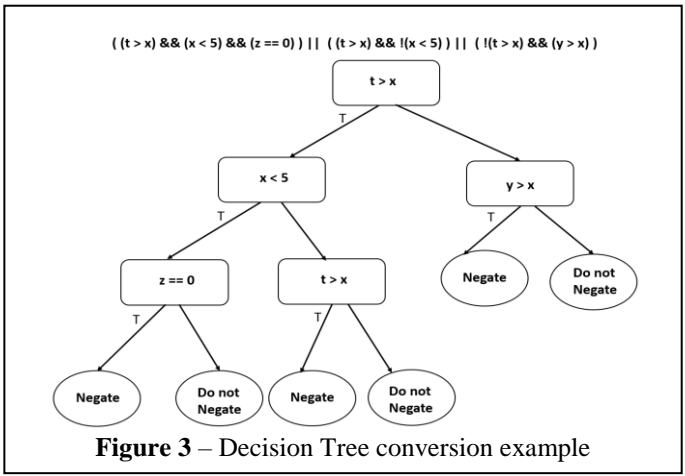
**Figure 3** – Decision Tree conversion example

4) A guard is created just before the candidate predicate. It executes an alternative *if* bytecode with flipped targets when the condition is met. The guard is identical to the one presented in Figure 4.

*E. Implementation*

Our implementation targeted the Java platform at the byte code level. Part of the work that posed most implementation challenges included *Heuristic CFA Search*, and *Training Data Collection* that both involved instrumenting and profiling Java byte code using the ASM Java bytecode manipulation and analysis framework (*asm.ow2.org*).

*F. CBFL: Identifying the Suspicious Predicates*

*CFAAR* requires a small number of suspicious predicates to be first identified, which could be achieved using some existing fault localization technique. However, since our experiments involved small programs, we opted to consider all predicates as suspicious, and thus used all of them as candidates for repairs. In future work we intend to devise an accurate CBFL technique suitable for *CFAAR*.

### III. EMPIRICAL EVALUATION

This section tries to answer the following research questions:

1) *RQ1: How Prevalent are the Defects that are Potentially Repairable by CFAAR?*
2) *RQ2: How Effective is CFAAR at Synthesizing Plausible Patches?*
3) *RQ3: How Effective is CFAAR at Synthesizing Correct Patches?*

In order to address these questions we applied our toolset to 149 single-fault subjects derived from 13 Java programs. Next, we describe the used subject programs then present and discuss our results.

*A. Subject Programs and Test Suites*

Our experiments involved 57 defective versions from the *Siemens* benchmark (*sir.unl.edu*) and 91 versions from the *Introclass* benchmark [19]. The *Siemens* subjects, namely, 8 *print_tokens2* versions, 4 *print_tokens* versions, 6 *replace* versions, 4 *schedule* versions, 1 *schedule2* version, 18 *tcas* versions, and 17 *tot_info* versions were manually converted to Java in [1]. Note that we excluded irrelevant bugs, those that could not be converted from C to Java, or those whose Java versions did not fail or caused exceptions to be thrown. The *Introclass* benchmark is originally written in C, it contains 6 programs (*digits*, *grade*, *median*, *smallest*, *syllables*, and *checksum*) and hundreds of related bugs. We opted to randomly select 20 versions from each program and convert them to Java. As a result, we used 20 *digits* versions, 20 *grade* versions, 20 *median* versions, 20 *smallest* versions, 4 *syllables* versions, and 7 *checksum* versions, for a total of 91 versions. Table 1 summarizes the information regarding the defective versions we used in addition to the test suite sizes. Note that the original test suites for the *Introclass* programs are very small; therefore, we randomly generated an additional larger test suite for each, referred to as $T_{large}$ in Table 1. However, some versions did not fail using $T_{large}$, for those we additionally used the original smaller test suite, denoted by $T$ in the table.

**Table 1-** Information about Subject Programs

| | | #Versions | |T| | |$T_{large}$| | LOC |
|---|---|---|---|---|---|
| *Siemens* | print_tokens | 4 | 4070 | - | 536 |
| | print_tokens2 | 8 | 4055 | - | 387 |
| | replace | 6 | 2843 | - | 554 |
| | schedule | 4 | 2650 | - | 425 |
| | schedule2 | 1 | 2710 | - | 441 |
| | tot_info | 17 | 1052 | - | 494 |
| | tcas | 18 | 1597 | - | 136 |
| *IntroClass* | digits | 20 | 16 | 1000 | 15 |
| | grade | 20 | 18 | 1000 | 19 |
| | median | 20 | 13 | 1000 | 24 |
| | smallest | 20 | 16 | 1000 | 20 |
| | syllables | 4 | 16 | 1000 | 23 |
| | checksum | 7 | 16 | 1000 | 13 |

*B. RQ1: How Prevalent are the Defects that are Potentially Repairable by CFAAR?*

A defect that is potentially repairable by *CFAAR* is one that could be fixed by negating one of its predicate statements at some instances during execution. In the context of our work, in order to get an estimated answer, we will assume that it is any defect for which **HeuristicCFASearch** makes one or more failing test exhibit correct behavior. Clearly, this is not a very accurate estimate since (currently) **HeuristicCFASearch** only explores a limited number of patterns, and only considers one predicate at a time as opposed to combinations of predicates.

The third row in Table 2 shows for each benchmark the number of the versions for which **HeuristicCFASearch** made some or all failing test cases behave correctly. The fourth row shows the numbers for which **HeuristicCFASearch** made all failing test cases behave correctly. On average, 58% of the versions had some or all of their failing test cases pass, and 30% had all of them pass. These findings suggest that the applicability of *CFAAR* is not very narrow.

**Table 2** – Summary of Results

| | Siemens | IntroClass |
|---|---|---|
| #Versions | 58 | 91 |
| Partially/Fully Fixed by **HeuristicCFASearch** | 28 | 59 |
| Fully Fixed by **HeuristicCFASearch** | 20 | 25 |
| Partially Fixed by Classifiers | 28 | 59 |
| Versions with Plausible Patches (fully fixed by some classifiers) | 24 | 17 |
| Versions with Correct Patches | 8 | 4 |

## C. RQ2: How Effective is CFAAR at Synthesizing Plausible Patches?

Whenever a given classifier was successful at making all failing test cases behave correctly, *CFAAR* will synthesize a corresponding plausible patch. Recall that a *plausible patch* is one that makes all the test cases pass (including those that were failing before the patch). Note that in many cases, multiple plausible patches could be generated for each defect, which calls for ranking them *w.r.t* likelihood of correctness. The fifth row in Table 2 shows for each benchmark the number of the versions for which the classifiers fixed some or all of the failing test cases. The sixth row shows the numbers for which some classifiers fixed all of the failing test cases. The numbers shown in the sixth row also represent the number of versions with plausible patches synthesized by *CFAAR*. Therefore, *CFAAR* was successful at generating plausible patches for 41 out of the 149 defects (i.e., 27.5%).

Next, we illustrate *Patch Synthesis* using patches for three versions used in our study, namely, *syllables v1*, *grade v13*, and *tcas v1*.

***Example 1 -*** The original code for *syllables v1* is faulty as it fails to check whether ch is equal to i:

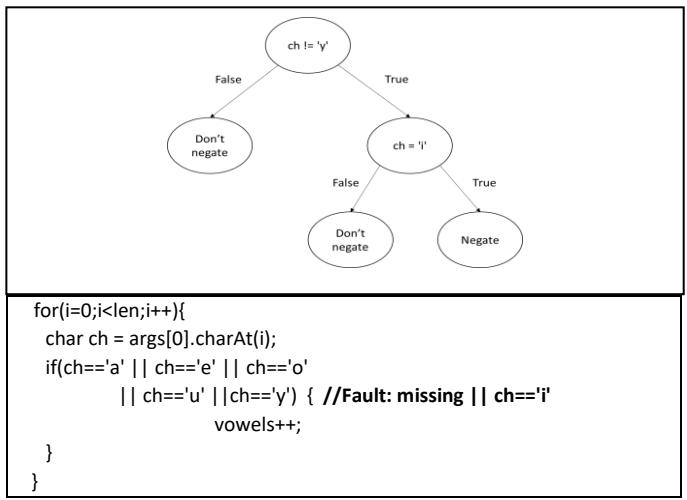

```
for(i=0;i<len;i++){
  char ch = args[0].charAt(i);
  if(ch=='a' || ch=='e' || ch=='o'
       || ch=='u' ||ch=='y') { //Fault: missing || ch=='i'
             vowels++;
  }
}
```

Shown above is a decision tree associated with one of the plausible patches for *syllables v1*.
The synthesized patch suggests replacing ch=='y' with:

```
ch=='y' ^ ((ch!='y' && ch=='i'))    // where ^ is xor
```

This plausible patch happened to be correct as it can be shown that it is semantically equivalent to the real fix:

```
ch=='y' ^ ((ch!='y' && ch=='i'))
  ⇔  ch=='y' ^ ch=='i'
  ⇔  (ch!='i' && ch=='y' || ch=='i' && ch!='y')
  ⇔  (ch='y' || ch='i')
```

***Example 2 -*** The original buggy code for *grade v13* is faulty because it mistakenly checks if the score is greater than *a* and *b*, instead of between *a* and *b*, as shown next:

```
if (score >= a){
    result += 'A';
}
else if ((score >= b) && (score > a)){ /* Fault –
           Potential fixes: 1) remove (score > a)
                            2) replace with (score < a) or (score <= a) */
    result += 'B';
}
else if ((score >= c) && (score < b)){
 ..
```

The decision tree consists of a single "negate" leaf node to be applied on the clause (score > a). Therefore, the synthesized patch suggests replacing (score > a) with !(score > a) or (score <= a).

***Example 3 –*** The previous two example patches happened to be correct. Applying *CFAAR* on *tcas v1* below yields a plausible patch that is actually incorrect:

```
result = !(Own_Below_Threat()) || ((Own_Below_Threat()) &&
(!(Down_Separation > ALIM()))); //fault: should have >=ALIM
```

Given the test suite associated with *tcas*, negating the faulty condition all the time was enough to make all test cases pass, which is clearly not a correct fix.

To improve the quality of our fixes, complementary approaches can be considered:

1) Improving the test suite by having more test cases cover the suspicious condition to help fine-tune the generated patches.
2) Ranking and prioritizing patches by looking at features such as syntactic/semantic distance to the faulty code, and similarity with documentation [40] and previous fixes [2].

## D. RQ3: How Effective is CFAAR at Synthesizing Correct Patches?

In order to assess our confidence in the correctness of the patches synthesized by *CFAAR* we followed two approaches. In case of *IntroClass*, we tested the patched subjects using validation test suites that we generated. The validation tests were programmatically created (rather easily) by generating random inputs. Out of the 17 plausible patches in *IntroClass*, 13 failed. That is, we have high confidence that 4 of our synthesized patches are correct.

Concerning the 20 *Siemens* plausible patches, using validation test suites was not feasible since it is hard to generate additional tests for these programs (noting that we used all existing tests for training). In this case, we opted to select a subset of the plausible patches to examine manually. The subset included 7 patches, of which we believe that 3 are correct and 4 are incorrect.

Table 3 compares our results to reported ACS, JGenProg, and Nopol results [40].

**Table 3** – Comparative Results

|          | *ACS* | *JGenProg* | *Nopol* | *CFAAR* |
|----------|-------|------------|---------|---------|
| Defects  | 224   | 224        | 224     | 149     |
| Plausible| 23    | 27         | 35      | 41      |
| Correct  | 18    | 5          | 5       | 12      |
| Incorrect| 5     | 22         | 30      | 29      |
| Precision| 78.3% | 18.5%      | 14.3%   | 29.3%   |
| Recall   | 8.0%  | 2.2%       | 2.2%    | 8%      |

*E. Threats to Validity*

A major threat to the external validity of our approach is the fact that our experiments involved a limited number of subject programs and faults. This could be remedied by conducting further experiments involving a variety of other subject programs from different domains and environments containing real and/or seeded defects.

In regard to threats to the internal validity of our approach, in its current state, *CFAAR* targets a rather narrow category of defects. However, we believe that the basic approach behind *CFAAR* is extendable to address defects that are repairable by a variety of alterations to a program's control flow and even data flow; which we intend to address in future work. Here we are referring to the methods adopted by *CFAAR*: a) to heuristically search for instances for when an alteration should be applied; b) to build classifiers based on state information; and c) to synthesize patches based on the classifiers.

## IV. RELATED WORK

Zhang et al. [42] presented a fault localization technique that is very relevant to our patch generation approach. It entails switching the valuation of the program's predicates, each one at a time for the purpose of producing the correct behavior. A predicate switch that yields a successful program completion can be further analyzed in order to identify the cause of the defect. Our approach differs in that: 1) due to our accurate CBFL technique, only few predicates need to be explored for switching; 2) predicate switching is considered at execution instances discovered by our approach; and most importantly, 3) a code patch is provided.

SPR [24] performs staged program repair. It performs fault localization using frequency analysis of positive and negative test case coverage. It leverages a set of parameterized transformation schemas (PTS) each of which targets a class of defects. For each PTS, SPR searches for an evaluation of schema parameters that allow the schema to produce a successful repair. It dismisses the transformation schemas (and their repairs) that fail the target value search and proceeds. For example PTS "if(cond || abstract_cond())" refines predicate "if(cond)" and "abstract_cond()" can return either *true* or *false*. If both target values do not fix the defect, the PTS dismisses. The last stage is condition synthesis where SPR uses the constraints obtained from the target value search to synthesize a condition. In particular, SPR selects states of program variables that are invariant for positive test cases (PPred), and states of program variables that were invariant for negative test cases (NPred). The latter are invariants that hold while the target value succeeds at fixing program behavior. The PTS synthesizes the condition such that the target values are obtained when NPred hold and PPred don't – "(!PPred) implies NPred". *CFAAR* is similar to SPR in that it uses both positive and negative test cases to synthesize a fixing condition. However, SPR performs a search for fixes matching existing schemas and consequently it has to use the values to determine the search rather than guide the search. *CFAAR* uses a classifier to determine whether a predicate needs to be changed and then uses the successful sequence of modified values to deploy a dynamic fix and synthesize a patch.

Precise condition synthesis [40] presents ACS to solve the overfitting problem in automatic defect repair. ACS splits the problem into (1) selecting the variables to be included in the precise condition, and (2) selecting the predicate from a set of existing predicates. It uses dependency order and analysis of API comments provided in javadocs to rank and select the variables, and uses predicate frequency in similar contexts to rank and select the predicates. *CFAAR* shares with ACS that it looks for more precise predicates to solve the overfitting problem. However, *CFAAR* inspects an infinite possible search space and uses tests to guide the search while ACS is restricted to existing code fragments.

Genesis [22] infers new patches from existing patches to fix (1) null pointer, (2) out of bounds, and (3) class cast defects. In a sense it refines the search space of [17] and [23] to concentrate on successful human patches instead of general code fragments, then it expands the potential search space by inferring transforms that generate defect patches from the existing patches. A transform is specified with two abstract syntax trees; one matches the faulty fragment in the original code, and the second specifies the replacement code. Generators allow introducing new logic and design elements in the fix specifically for template variables that are not matched in the code. Integer linear programming (ILP) is used to limit the search space to a reasonable number of patches by maximizing the number of training patches that cover the inferred search space. Unlike ACS and SSCR, Genesis is not limited to existing code fragments. However, it is limited to patches that are syntactically related to the existing patches through an AST. *CFAAR* differs in that it inspects a search space that is related semantically to the defect and uses the test cases to guide the search.

Semantic search for code repairs (SSCR) [17] characterizes a large set of code fragments with FOL constraints and considers those potential fixes, relates a fault in a program to fragments of code in the program and characterizes these fragments with fault FOL constraints, then it uses constrain solvers to match the fault constraints with the potential fix constraints. The technique finally integrates the selected fix by syntactic modifications such as renaming variables. *CFAAR* is similar in the sense that it also performs a semantic search, it differs in the sense that it is not limited to a large set of existing code fragments. *CFAAR* searches an infinite space of potential fixes that is the composition of modifications of failing control statements and uses test cases to guide the search. While the work in [17] relies heavily on computationally expensive SMT solvers, *CFAAR* leverages decision tree algorithms as heuristics to construct the fix.

Le Goues et al. [18] proposed *GenProg*, a repair technique based on genetic programming. They assume that repairing a fault in one function can make use of snippets of code appearing in other functions in the program. For example, several existing functions in a program might implement checks for whether a pointer is null, the corresponding code can then be inserted in the function under repair in the aim of repairing it. The technique explores different variations of the defective program such as those resulting from inserting statements, deleting statements, and swapping statements. Also, mutation and crossover operators are applied and guided using a fitness function that evaluates the generated program against the test suite. Once a repair is found, it is further refined using delta debugging by discarding the unnecessary statements within. Our repair technique is very different in terms of its underlying approach and the nature of the produced solution.

Assiri and Bieman [4] evaluated the impact of ten existing CBFL techniques on program repair. Specifically, they measured their impact on the effectiveness, performance, and repair correctness of a brute force program repair tool, i.e., a tool that exhaustively applies all possible changes to the program until a repair is found. A brute-force repair tool is guaranteed to fix a fault if a repair is feasible. Therefore, a failure to find a potential repair would likely be related to the selected CBFL technique. Including our proposed CBFL technique in their comparative evaluation would be valuable, as it could help justify its cost.

Martinez and Monperrus [25] presented *Astor*, a library comprising the implementation of three major program repair approaches for the Java platform. The library is also meant to be extended by the research community by adding new repair operators and approaches. The currently supported approaches that originally targeted C programs are:

1) *jGenProg2*: an implementation of *GenProg* for Java [37][18] in which repair operators only consider nearby code, and not the whole codebase as it is the case in *GenProg*.
2) *jKali*: an implementation of the *Kali* approach [32] for Java, which performs repair by exhaustively removing statements, inserting return statements, and switching predicates. Our approach is far from being exhaustive since the predicate switching is highly targeted in terms of location and time.
3) *jMutRepair*: an implementation of the approach presented by Debroy and Wong [8] for the java platform. *jMutRepair* mutates the relational and logic operators in suspicious if condition statements. Since our approach negates predicates at the byte code level (single clause predicates), it practically also mutates relational and logic operators. However, unlike *jMutRepair*, our approach negates the predicates at specific execution instances.

Nopol [9][41] uses angelic fix localization to locate faulty if-then-else conditions, execute passing test cases to compute a model of the correct behavior of the program, abstract the values of the variables in the model to FOL constraints, and uses SMT solvers to compute a fix to the condition. The technique targets buggy *if* conditions and missing preconditions. *CFAAR* is similar to Nopol in that it uses both the passing and the failing test cases to compute a model for the fix. *CFAAR* differs in the abstraction techniques as it is variable specific while Nopol creates SMT statements to model the execution. Finally, *CFAAR* uses decision trees to compute a potential code fix, while Nopol uses SMT solvers.

An influencing precursor of *Nopol* is *SemFix*, an approach presented by Nguyen et al. [28]. *SemFix* is based on symbolic execution, constraint solving, and program synthesis. Given a candidate repair location $l$, *SemFix* derives constraints on the expression to be injected at $l$ in order to make the failing test case pass. Symbolic execution is used to generate the repair constraints, and program synthesis is used to generate the repair patch. Similar to *SemFix*, *DirectFix* [27] and *Angelix* [26] also aim at synthesizing repairs using symbolic execution and constraint solving; but are more scalable.

Tan and Roychoudhury [36] presented *relifix* an approach for repairing regression bugs. The mutation operators considered are derived by manually inspecting real regressions bugs. The potential repair locations were identified by differencing the current version of the defective program with its previous version, and by considering the *Ochiai* suspiciousness of the locations.

Pei et al. [29] proposed an approach that exploits contracts such as pre/post-conditions to localize faults and generate repairs in Eiffel programs.

Elkarablieh and Khurshid [12] developed a tool called *Juzi*, within which the user provides a Boolean function that checks whether a given data structure is in a *good* state. The function is invoked at runtime, and in case a corrupt state is detected, the tool performs repair actions via symbolic execution. One of the authors later targeted the repair of the selection conditions in SQL *select* statements [13].

V. CONCLUSIONS

We presented *CFAAR*, a test-based repair assistance technique that targets defects that are repairable by *negating* control statements under some specific conditions. *CFAAR* relies on the program's state to determine when a candidate control statement should be negated in order to yield correct behavior. A synthesized patch is generated based on the state information, in the form of a conditional that guards the candidate control statement. When presented with a patch, the developer would: 1) use it as is, if deemed correct; or 2) use it as assistance during the debugging process.

Our experiments involving 149 defects revealed the following: 1) 91 defects were found to be potentially repairable by *CFAAR*; 2) 41 plausible patches were generated by *CFAAR*; and 3) at least 12 plausible patches are believed to be correct. In the future, we will conduct experiments aiming at assessing the level of repair assistance our plausible patches provide to the developer.

ACKNOWLEDGMENT

This research was supported in part by the Lebanese National Council for Scientific Research, and by the University Research Board at the American University of Beirut.